\documentclass[accepted=2019-05-24]{quantumarticle}

\usepackage{amsmath,amssymb}
\usepackage[]{graphicx}
\usepackage[utf8]{inputenc}
\usepackage[colorlinks,linkcolor=blue,citecolor=blue,urlcolor=blue,breaklinks=true]{hyperref}
\usepackage{cleveref}
\usepackage{soul}

 \newcommand{\comm}[2]{\left[ \, #1 \, , \, #2 \, \right]}
 \newcommand{\bra}[1]{\left\langle #1 \right |}
 \newcommand{\ket}[1]{\left | #1 \right\rangle}
 
 \newcommand{\avr}[1]{\left\langle #1\right\rangle}

\graphicspath{{./figuresHard/}}

\begin{document}

\title[Finite-size scaling of a dissipative quantum phase transition]{Finite-size scaling of the photon-blockade breakdown dissipative quantum phase transition}
\author{A. Vukics}
\address{Wigner Research Centre for Physics, H-1525 Budapest, P.O. Box 49., Hungary}
\email{vukics.andras@wigner.mta.hu}
\orcid{0000-0001-8916-4033}
\author{A. Dombi}
\address{Wigner Research Centre for Physics, H-1525 Budapest, P.O. Box 49., Hungary}
\orcid{0000-0003-2158-0609}
\author{J. M. Fink}
\address{Institute of Science and Technology Austria, 3400 Klosterneuburg, Austria}
\orcid{0000-0001-8112-028X}
\author{P. Domokos}
\address{Wigner Research Centre for Physics, H-1525 Budapest, P.O. Box 49., Hungary}
\orcid{0000-0002-1002-5733}


\begin{abstract}
We prove that the observable telegraph signal accompanying the bistability in the photon-blockade-breakdown regime of the driven and lossy Jaynes–Cummings model is the finite-size precursor of what in the thermodynamic limit is a genuine first-order phase transition. We construct a finite-size scaling of the system parameters to a well-defined thermodynamic limit, in which the system remains the same microscopic system, but the telegraph signal becomes macroscopic both in its timescale and intensity. The  existence of such a finite-size scaling completes and justifies the classification of the photon-blockade-breakdown effect as a first-order dissipative quantum phase transition.
\end{abstract}


\maketitle

\section{Introduction}

A number of recent theoretical proposals and experiments refer to a new type of dissipative phase transition in a quantum system, that is said to be of first order. The corresponding systems include cluster formation of Rydberg-state atoms, described by Ising-type spin models \cite{Ates2012Dynamical,Carr2013Nonequilibrium,Marcuzzi2014Universal,Malossi2014Full,Overbeck2017Multicritical,Letscher2017Bistability}, the vacuum-superfluid transition in a dissipative Hubbard model \cite{Labouvie2016Bistability}, arrays of nonlinear cavity systems \cite{Leboite2013Steady}, and the ultimately simple model system of a single semiconductor cavity polariton mode \cite{Casteels2017Quantum,Casteels2017Critical,Rodriguez2017Probing,Fink2018Signatures}. Common to these systems is that the signature of the first-order transition is the bistable behaviour of an observable. Typically, the bistability is manifested by a telegraph-like signal, i.e.,  randomly alternating ‘bright’ and ‘dark’ periods of a radiation mode, associated with the mixture of two phases of the quantum system. This temporal signal can be readily monitored by a ‘classical’ detector. For example, this is the case in the photon-blockade breakdown transition \cite{carmichael2015,dombi2015bistability,palyi2012spin}, which has recently been experimentally demonstrated \cite{Fink2017Observation}. However, such a signal can be observed in other scenarios that are not regarded as phase transitions. Think of, for example, the famous quantum jump experiment \cite{Bergquist1986Observation}  measuring the fluorescence of a driven V-type three-level atom. The fluorescence signal also exhibits long bright and dark periods, separated by single quantum jumps. The ‘dark’ period indicates the atom being in a long-lived, metastable state, whereas a spontaneous emission event can take it into the ground state, where the atom is resonantly driven by a laser on an electric-dipole-allowed ‘bright’ transition. What is then the distinctive feature of a first-order phase transition? 

This paper is devoted to complete and justify the interpretation of the photon-blockade-breakdown effect as a first-order dissipative quantum phase transition. To this end, one needs to introduce the concepts of \emph{thermodynamic limit} and \emph{finite-size scaling} for the microscopic system of a driven dissipative Jaynes-Cummings model. The thermodynamic limit will be defined for this microscopic system in such a way that the number of relevant degrees of freedom remains fixed. Instead of growing the system size, we construct a scaling of the parameters which keeps the form of the stationary solution of the driven-dissipative system invariant. At the same time, the proposed finite-size scaling leads to an increasing robustness of the attractor states associated with the bistability signal, until these states reach full stability in the thermodynamic limit. On approaching this limit, at variance with the fluorescence shelving experiment, no \emph{single} quantum jump or other microscopic event can flip the system from one phase to the other. In the thermodynamic limit, the blinking telegraph-like signal vanishes completely and the state of the system is determined by the initial condition, similarly to the usual hysteresis behaviour in classical critical systems. If such a finite size scaling is possible – and here we show that this is the case for the photon-blockade breakdown effect –, then the bistability that can be observed in a given experimental realization of the system with its finite parameters not in the thermodynamic limit, can be considered the finite-size approximation of what is a genuine first-order phase transition in the thermodynamic limit.

The paper is structured as follows. In \cref{sec:QPTnotions} we clarify the basic notions of dynamical quantum phase transitions in dissipative systems, in particular, the concept of a first-order transition. In \cref{sec:PBBnutshell} we recall the model and the basic ingredients of the photon-blockade-breakdown effect, together with the equations governing the underlying classical phase transition. In \cref{sec:telegraph} we present the time evolution of the system in steady state, which we show to be a telegraph signal. We undertake a detailed numerical analysis of timescales and the statistics of dwell times in the attractors. In \cref{sec:fillingFactor}, the finite-size scaling of the parameters is determined, and \cref{sec:DwellTimes} is devoted to the  analysis of the stability of phases. We shortly comment on the role of atomic decay in \cref{sec:AtomicDecay} and then conclude in \cref{sec:conclusion}.

\section{What is a first-order dissipative quantum phase transition?}
\label{sec:QPTnotions}

Quantum phase transitions (QPTs) are abrupt, symmetry-breaking changes in the quantum state of a large quantum system at a critical value of an external control parameter. The QPT takes place at zero temperature and thus it generally refers to a change of the ground state \cite{Sachdev2011Quantum}. The ground states on the two sides of the critical point have different symmetries and correspond to different macroscopic observables when a quantum measurement is performed. The concept of QPT can be straightforwardly extended to dissipative (open) quantum systems. Assume that the system is excited by an external coherent field, e.g., by laser irradiation. The coherent driving competing with the various damping processes sets the system in a dynamical equilibrium which is not the ground state, moreover, it may be far from a thermal state due to energy flow through the system. Clearly, the steady state will depend on external parameters. Just like the ground state of closed systems, the stationary state of driven-dissipative quantum systems can also exhibit non-analytic behaviour as a function of some control parameter \cite{Marino2016Driven,Carmichael2018Dissipative,reiter2018cooperative}. All this can happen at zero temperature. The critical point is marked by diverging quantum fluctuations and corresponds to a symmetry breaking change of the \emph{steady-state} \cite{Nagy2011Critical,Brennecke2013Observation}. Note that the presence of dissipation in the open system amounts to information leakage to the environment. Therefore, if such a transition occurs in the quantum system, the different steady states under continuous measurement are associated with different measured, ‘classical’ outputs. In this sense the quantum states are ‘amplified’ to \emph{phases} with different macroscopic observables even in the case of a small system with only few degrees of freedom.

Besides continuous phase transitions, theoretical and experimental works pointed out recently that first-order phase transitions can also have their counterpart in open quantum systems  \cite{carmichael2015,Fink2017Observation}. First-order phase transitions are benchmarked by the \emph{coexistence of phases}. Instead of a critical point separating two phases, there is a critical region in which bistability and hysteresis occur \cite{Bonifacio1978}. How can the coexistence of ‘macroscopically distinct’ states be accommodated into quantum theory? The state of a driven-dissipative system is expressed by a density operator. The steady-state density operator, which obeys a linear algebraic equation, must be a continuous function of the external control parameter of the system. However, the density operator can represent a bimodal distribution, a mixture of two macroscopically distinct components. There can be then a ‘rift’ separating the two ‘phases’, meanwhile the transition of the density matrix  can occur smoothly from one phase to the other when tuning the control parameter. It is the weights in the mixture which vary continuously as a function of the control parameter: one of the components vanishes gradually with the simultaneous growing of the weight of the other. The complete transition can happen in a \emph{finite range of the control parameter}, which is the bistability range.  

The starting point of our study is that the above picture of 1st order QPT is, however, not the full story. In particular, the bimodal steady-state density operator does not tell us about the stability of the phases. It allows only for identifying the two phases and their relative weights. In the bistability range, the system continuously monitored in its dissipation channel manifests a randomly blinking signal between the distinct values associated with the two phases. The same steady-state density operator can correspond to different characteristic dwelling times in the blinking signal. This is the goal of unraveling into quantum trajectories, and performing a finite-size upscaling in the language of these trajectories. The steady-state density matrix is kept invariant under the upscaling of the system parameters, only with increasing intensity in the bright periods, meanwhile the stability of the phases increases, that is, the characteristic dwell time in the attractors tends to infinity. In the following, we will perform this analysis in the case of the photon-blockade-breakdown phase transition.

\begin{figure}
\includegraphics[width=\linewidth]{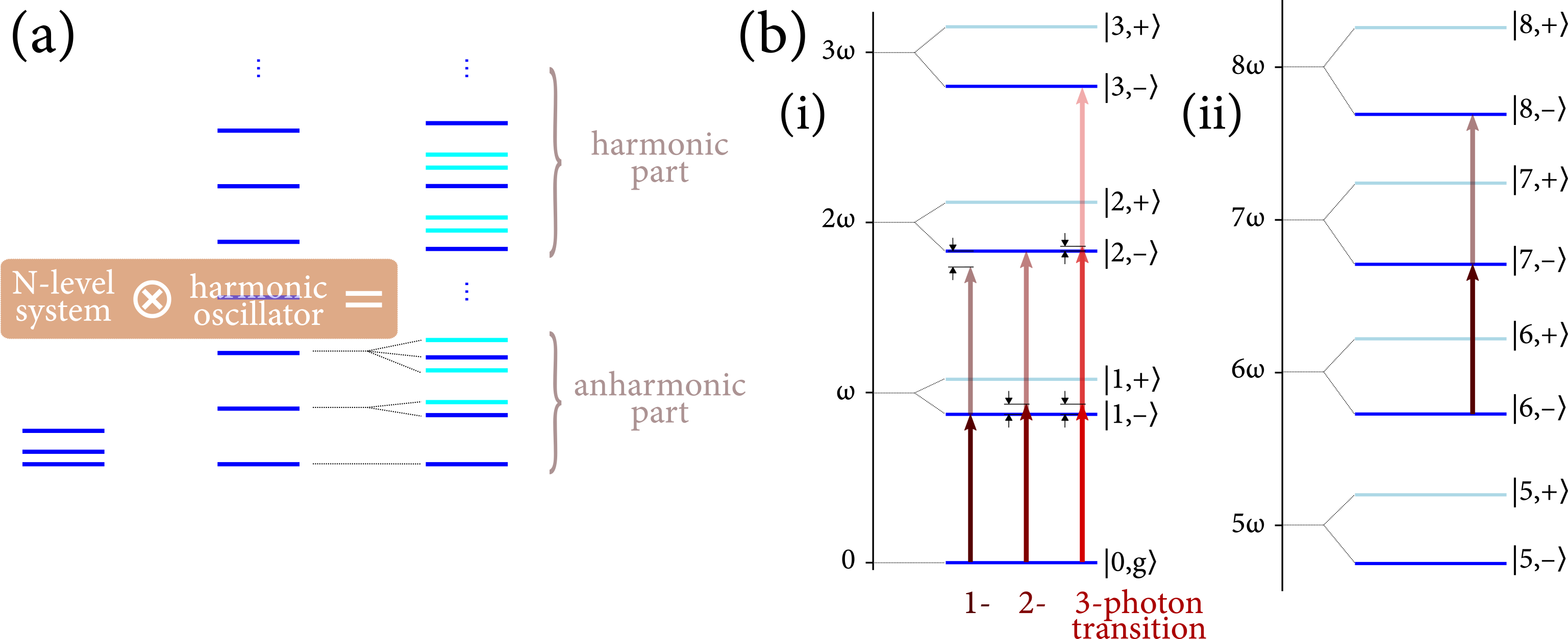}
 \caption{(a) In a quantum system consisting of a finite-level system strongly coupled to a harmonic oscillator, the lower part of the spectrum is anharmonic, but there are always higher-lying harmonic subsets of the spectrum. (b) The Jaynes-Cummings spectrum is a prototype of this behaviour (this part of the figure is to scale).  Panel (i): the anharmonic part of the spectrum. An excitation tuned to a 1-photon transition misses the second rung of the ladder by a significant amount. That tuned to a 2- and 3-photon transition, misses the first rung and the first and second rung, respectively. Panel (ii): closely harmonic part of the spectrum. An excitation tuned to the transition between $\ket{6,-}$ and $\ket{7,-}$ is close to resonance also with the transition between $\ket{7,-}$ and $\ket{8,-}$ (the detuning is invisible on the figure’s scale).}
 \label{fig:coherentIncoherent}
\end{figure}

\section{The photon-blockade-breakdown phase transition in a nutshell}
 \label{sec:PBBnutshell}

\subsection{The driven-lossy Jaynes--Cummings model}

The simplest possible system exhibiting the photon-blockade-breakdown effect is composed of a two-level system coupled to a harmonic oscillator. The two-level system can be an atom or artificial atom, whereas the oscillator can represent a single lossy mode of the radiation field or a longitudinal mode of a stripline resonator. We describe this interaction within the driven Jaynes-Cummings model, i.e., using the electric-dipole coupling and the rotating-wave approximation (RWA):
\begin{multline}
\label{eq:H_time}
H=\omega_\mathrm{M}\,a^\dagger a\,+\,\omega_\mathrm{A}\, \sigma^\dagger\,\sigma\, +\, i g \left( a^\dagger\,\sigma-\sigma^\dagger\,a\right)\\+\,i \eta\left( a^\dagger\,e^{-i\omega t}-a\,e^{i\omega t}\right)\; ,
\end{multline}
where $\omega_\mathrm{M}$ is the angular frequency of the mode with boson operator $a$, $\omega_\mathrm{A}$ is that of the atomic transition with step-down operator $\sigma$, and $g$ is the single-photon Rabi frequency expressing the coupling strength. The external coherent drive is characterized by the amplitude $\eta$, here taken to be real for simplicity, and the frequency $\omega$. We note that due to the strong coupling, our treatment is close to the validity limit of the RWA. A quantum-to-classical transition in a similar system without the conventional rotating-wave approximation has been reported in Refs. \cite{pietikainen2017,pietikainen2019}.

Assuming resonance between the mode and the atom, i.e. $\omega_\mathrm{M} = \omega_\mathrm{A}$, and going into a frame rotating at the drive frequency, one gets a virtually time-independent Hamiltonian,
\begin{equation}
\label{eq:H}
H=-\Delta\,\left(a^\dagger a + \sigma^\dagger\,\sigma\right) + i g \left( a^\dagger\,\sigma-\sigma^\dagger\,a\right) +\,i \eta\left( a^\dagger-a\right)\; ,
\end{equation}
where the detuning $\Delta=\omega-\omega_M$ is a tunable parameter of the drive. The mode is that of a high-finesse resonator and is subject to loss. Similarly, the two-level system can have decay through spontaneous emission. These incoherent processes can be modelled by Liouvillian terms in the master equation 
 \begin{multline}
 \label{eq:master}
	\dot{\rho} = -i\comm{H}{\rho} + \kappa\left( 2 a \rho a^\dagger - a^\dagger a \rho - \rho a^\dagger a \right)\\ + \gamma \left( 2 \sigma \rho \sigma^\dagger - \sigma^\dagger \sigma \rho - \rho \sigma^\dagger \sigma  \right)\;.
\end{multline}
In the rest of the paper we will consider the case $\kappa\gg\gamma$, most importantly $\gamma =0$. The mode relaxation parameter $\kappa$ defines the microscopic timescale of the problem. 

\subsection{The photon-blockade-breakdown effect}

For weak drive strengths $\eta\ll g$, the excited eigenstates of the Hamiltonian are close to the Jaynes-Cummings dressed states
\begin{equation}
\ket{n,\pm} =\frac{1}{\sqrt 2}\left(\ket{g,n} \pm \ket{e,n-1} \right) 
\end{equation}
with $n=1,2,\dots$, and the energy levels are
\begin{equation}
E_{n,\pm}=n\,\omega_\mathrm{M}\pm \sqrt{n} g\;. 
\end{equation}
In the strong coupling regime $g\gg \kappa$, the level shifts $\pm \sqrt{n}\, g$ with respect to the bare frequencies exceed significantly the linewidth $\sim \kappa$, so the system cannot be excited out of the ground state $\ket{g,0}$ by a near-resonant driving $\Delta\approx 0$. This is the photon blockade effect.\footnote{In the literature, the ‘photon blockade’ sometimes denotes the effect when the first excited state with a single photon can be excited resonantly, but further excitations are suppressed due to off-resonance. This is analogous to the effect of Coulomb blockade to some extent. Here we use the term in a more general sense, where the system is blocked in the ground state, and no photons at all can be transferred to the system.}

Above a certain intensity of the driving, however, the system can be excited via higher-order (multi-photon) transition processes into the higher-lying part of the spectrum. Since one of the constituents is a harmonic oscillator, in the high-lying part there are harmonic subsets of the spectrum (cf. \cref{fig:coherentIncoherent}), ladders of equidistant levels which can host a coherent state with large amplitude and well-defined phase. Following a low-probability multi-photon transition into this range of the spectrum, the system gets stabilized into such a semiclassical trapping state by the competition of coherent driving and decay. Such an excitation, i.e., \emph{the breakdown of the photon blockade} takes place in the form of a bistability. In a finite interval of the drive strength, the steady-state density operator of the system is the mixture of the ‘dark’ and ‘bright’ phases, i.e., the ground state and a highly excited coherent state of the oscillator \cite{carmichael2015,dombi2015bistability}. As a main subject of our study,  we will analyse this solution of the density matrix in detail starting from the next section. However, beforehand, we try to capture the non-analytic behaviour in a corresponding classical theory.

\subsection{Classical phase diagram}

\begin{figure}
\centering \includegraphics[width=\linewidth]{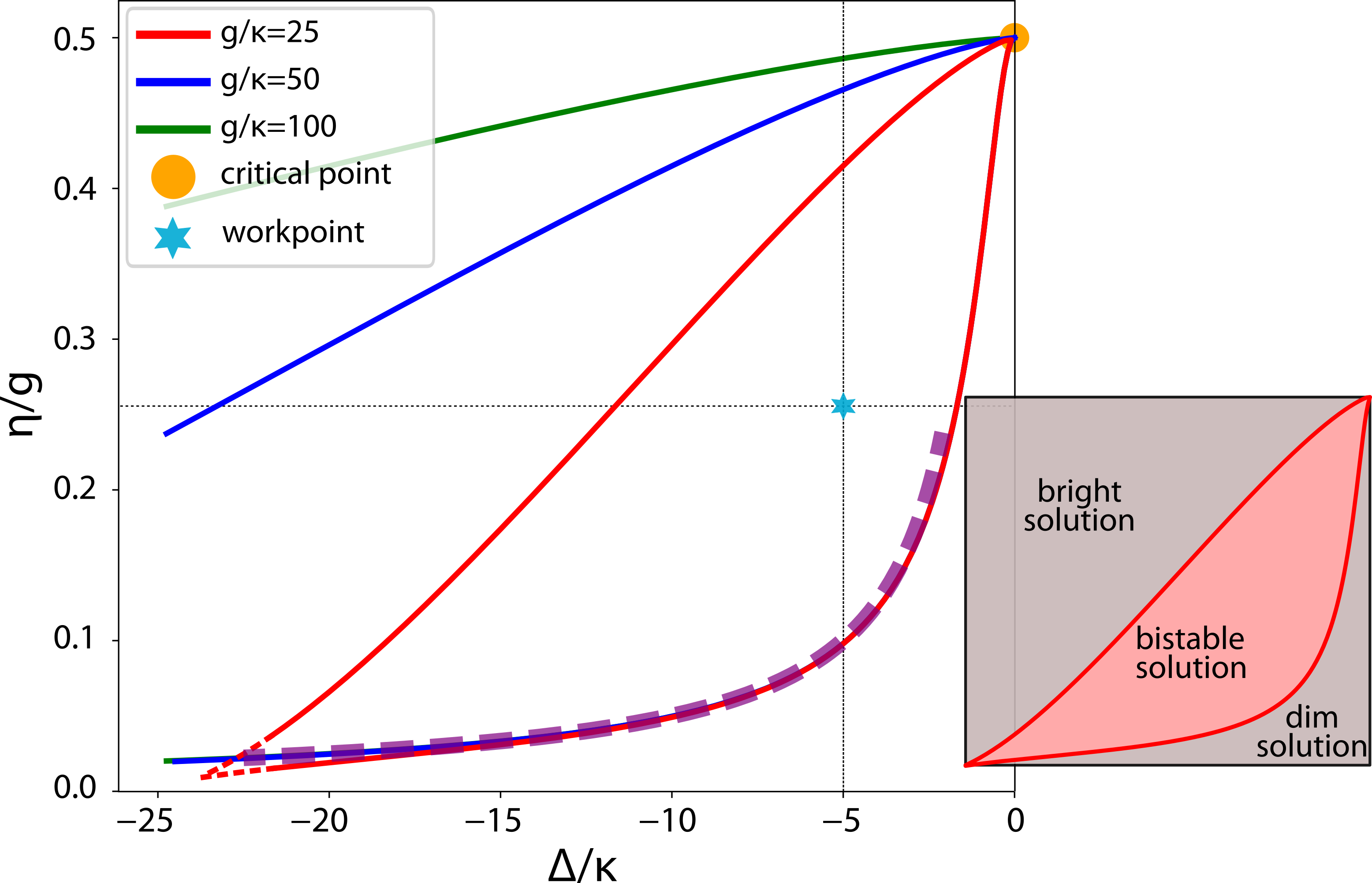}
\caption{Phase diagram of the neoclassical theory based on a numerical solution of the transcendental \cref{eq:nscale} for three different values of the coupling constant $g$. The low boundaries closely overlap, the upper boundaries differ for the three values of $g$. As the numerics becomes very unstable when approaching the far-off-resonance closing point of the bistability region, the closing of the red curves on the left is inferred by extrapolation (dashed segments). The common closing point on the right for all $g$ values is the spontaneous dressed-state polarization critical point. The cyan star denotes the workpoint chosen in this paper: it is at this detuning and around this drive strength that we are going to study the bistable solution for several $g$ values. Magenta dashed line: the inference \labelcref{eq:eta_low} for the lower limit of the bistability region.}
\label{fig:PhaseDiagram}
\end{figure}

Following Carmichael \cite{carmichael2015}, let us first look at what the Jaynes and Cummings semiclassical (also known as ‘neoclassical’) equations can tell us. These are obtained by taking expectation values in Heisenberg equations of motion (e.g. $\alpha=\avr{a}$) and factorizing the expectations of operator products to obtain a closed set of nonlinear equations for the expectation values. Taking $\gamma=0$, the theory leads to the self-consistent equation (valid for $\Delta<0$)
\begin{multline}
\label{eq:nscale}
\frac{|\alpha|^2}{N_{\rm scale}} = \left(\frac{2\eta}{g}\right)^2 \\\times \left[ 1 + \left( \frac{\Delta}{\kappa} - \frac{1}{\sqrt{\Delta^2 \kappa^2/{g^4} + {|\alpha|^2}/{N_{\rm scale}} }}\right)^2\right]^{-1} \; ,
\end{multline}
where we introduced the parameter $N_{\rm scale} = g^2/4 \kappa^2$. This nonlinear equation can afford different solution sets. The various domains are depicted in the phase diagram in \cref{fig:PhaseDiagram}. Below the lower boundary, the phase is the photon blockade regime, whereas above the higher boundary, the system is highly excited. In between, \cref{eq:nscale} has multivalued solution indicating bistability. The coordinates were chosen to be the tunable variables, i.e., the frequency and amplitude of the driving, which can serve as control parameters of the phase transition. Only the $\Delta<0$ half-plane is shown, the positive-detuning part being the same mirrored to the $\Delta=0$ axis.

The neoclassical result suggests an appropriate thermodynamic limit \cite{carmichael2015} and a corresponding scaling of system parameters. On fixing the timescale to the microscopic one, $\kappa=1$, we see that a characteristic photon number is expected to scale as $\sim  g^2$. Hence, the thermodynamic (infinite-system) limit is the strong coupling limit $g\to\infty$ (in contrast to previously-reported thermodynamic or classical limiting cases of quantum phase transitions in the Jaynes-Cummings and Rabi models \cite{Hwang2015,Hwang2016,Larson2017Superradiant,Hwang2018}). Simultaneously, the drive amplitude must also be scaled up. The first guess, cf.~the prefactor on the right hand side of  eq.~(\ref{eq:nscale}), would be $\eta\rightarrow\infty$ such that $\eta/ g$ is kept invariant. This is why the quantity $\eta/g$ is used for the drive amplitude on the vertical axis of the  phase diagram. With this scaling, the lower boundary of the bistability phase becomes indeed independent of $g$: the three curves coincide perfectly. On the other hand, the upper boundaries reveal a further dependence on $g$. That is, the solution sets of the neoclassical equations are not invariant under the transformation of $g,\eta\rightarrow\infty$ with $\eta/ g = \mathrm{const}$. Later, in sec.~\ref{sec:fillingFactor} we will identify the non-trivial scaling of $\eta$ which leads to a properly defined thermodynamic limit of the system.

Nevertheless, since the lower boundary of the neoclassical bistability domain is invariant, and also the upper boundary does not vary strongly at $\Delta=-5\kappa$ for the g values shown in \cref{fig:PhaseDiagram} and used in this paper, in the plots the drive strength $\eta$ is given in units of $g$. 

Finally, let us make two side remarks. Firstly,  Gutiérrez-Jáuregui and Carmichael \cite{gutierrez2018dissipative} pointed out that another possible scaling deducible from the mean-field steady state equations is keeping \(\Delta/g=\mathrm{const.}\) This would make that in the phase diagram \cref{fig:PhaseDiagram}, the upper limiting curves of the bistability domain would coincide, but the lower ones would differ for different \(g\) values. So, at this point the two scalings \(\Delta/\kappa=\mathrm{const.}\) and \(\Delta/g=\mathrm{const.}\) appear as equivalent alternatives. It can be argued that our choice is more natural in the sense that the detuning from a (bare) resonance is measured against the width of that resonance.

Secondly, there is a critical point at $\eta/g=1/2$ specific to the resonant driving case $\Delta=0$, where the lower and upper limits of the bistability region converge. It separates the solution $\alpha=0$ with population inversion increasing from $-1/2$ to $0$ from the one with $\sigma_z=0$ and increasing $\alpha$ as the drive strength $\eta$ is increased further. This result is in accordance with that of the full quantum treatment which can be pursued to an analytical solution in the resonant case  \cite{alsing1992dynamic}. It shows that the quasienergies coalesce in this critical point. This critical behaviour was identified as the spontaneous dressed-state polarization by Alsing and Carmichael \cite{alsing1991spontaneous}. 

\subsection{Bistability in the quantum solution and intuitive explanations}

Numerical simulations of the full quantum problem defined by \cref{eq:master} confirm qualitatively the phase diagram based on the neoclassical equations. The existence of a bistability regime close to resonance $|\Delta| \ll g$ has been confirmed  \cite{carmichael2015,dombi2015bistability}. The stationary density-operator solution of the master equation is then a statistical mixture of two states: the ‘dim’ state where the field is close to the vacuum and the atom is in the ground state, and a ‘bright’ state which consists of a highly excited coherent state of the field (and a completely saturated atom). On increasing the drive strength $\eta$ at a fixed detuning, the relative weight of the two components is continuously varied such that the probability of the bright component goes from zero to 1 in a finite range of $\eta$ \cite{Fink2017Observation}. The steady state is hence a continuous function of the parameters, however,  there is a ‘rift’ between the two components of the mixture: these are classically discernible states.

The stability of the bright component for high-enough drive strength can be understood intuitively as a balance of coherent drive and spontaneous decay in a harmonic oscillator. The frequency separating adjacent dressed states $\ket{n+1,-}$ and $\ket{n,-}$ is $\omega_\mathrm{M} - g (\sqrt{n+1}-\sqrt{n}) \approx \omega_\mathrm{M} - g/(2\sqrt{n})$. In the limit of large photon numbers $n\to N$, this tends to a harmonic ladder in which the coherent drive $\eta$ and the decay $\kappa$ compete to create a coherent state with amplitude
\begin{equation}
\label{eq:resonance}
\alpha=\frac\eta{\kappa-i\left(\Delta+\frac{g}{2\sqrt{N}}\right)}\;\Longrightarrow\; N= \frac{\eta^2}{\kappa^2 + \left( \Delta + \frac{g}{2\sqrt{N}}\right)^2} \; ,
\end{equation}
where the self-consistent equation for the photon number $N$ was obtained by taking the absolute value squared of the amplitude $\alpha$. The smallest drive strength for which this equation can be satisfied is in the case of ‘resonance’, i.e. when the expression in the parentheses in the denominator vanishes: $\Delta=-{g}/{(2\sqrt{N})}$. The self-consistent solution is then $N=(\eta/\kappa)^2$, from which the minimum drive strength follows as
\begin{equation}
\label{eq:eta_low}
\frac{\eta_{\rm min}}{g} \simeq \frac{\kappa}{2|\Delta|}\;.
\end{equation}
This law is drawn in the thick dashed magenta line in the phase diagram in \cref{fig:PhaseDiagram}, and coincides quite accurately with the lower border of the classical bistability domain. It is remarkable that the solution of a classical equation obeys a law extracted from intuitive consideration of quantized energy levels. On  further increasing the drive strength, the photon number increases, thereby leading to some detuning $\Delta + {g}/{(2\sqrt{N})} \neq 0$ in \cref{eq:resonance}. However, there may still be a self-consistent solution. Because the lower part of the harmonic ladder is missing (that part of the spectrum being the anharmonic photon-blockading part), there is no deterministic evolution into such a self-consistent solution. Nevertheless, the displacement operator corresponding to coherent driving contains multi-photon excitation processes, so the excitation into the near-resonant regime with a self-consistent solution can take place in a probabilistic manner.

The upper limit of the bistability domain cannot be determined from an argument as simple as the above for the lower limit. The reason is that the more we increase the drive strength $\eta$, the more the quasi-energy levels, i.e. the true eigenvalues of the Hamiltonian (\ref{eq:H_time}), differ from the dressed levels of the $\eta=0$ Jaynes-Cummings model, since they are getting dressed also by the coherent drive \cite{alsing1992dynamic}. However, the analytical form of the quasi-energy levels in the finite-drive strength case is not known for $\Delta\neq 0$, only for the case of $\Delta=0$. Nevertheless, it is clear that the appearance of an $\eta$-dependence of the energy levels makes that the $\eta/g$ scaling suggested by both \cref{eq:nscale,eq:resonance} is disrupted for large $\eta$ values.


\section{The telegraph signal}
\label{sec:telegraph}

Not only the classical phase diagram, but even the steady-state density operator solution of the full quantum problem defined by \cref{eq:master} does not describe all the relevant aspects of our phase transition. In the following we prove by numerical simulation that the components of the mixture become robust classical attractors in the thermodynamic limit. To this end, we need to unravel the density operator into the time domain, so that we can extract the dwell timescales, we can show their divergence, and we can determine the relevant exponents. To this end, we use the quantum trajectories generated by the Monte-Carlo wavefunction method. In principle, the ensemble average of many trajectories yields a density operator evolving in time towards the steady-state one. However, in the ergodic case (which will be our assumption here), the temporal averaging of the stochastic state vectors along a single long quantum trajectory yields the same steady-state density operator. Therefore it makes sense to consider a trajectory as an actual evolution under continuous measurement with an ideal photodetector.

\begin{figure*}
 \includegraphics[width=\linewidth]{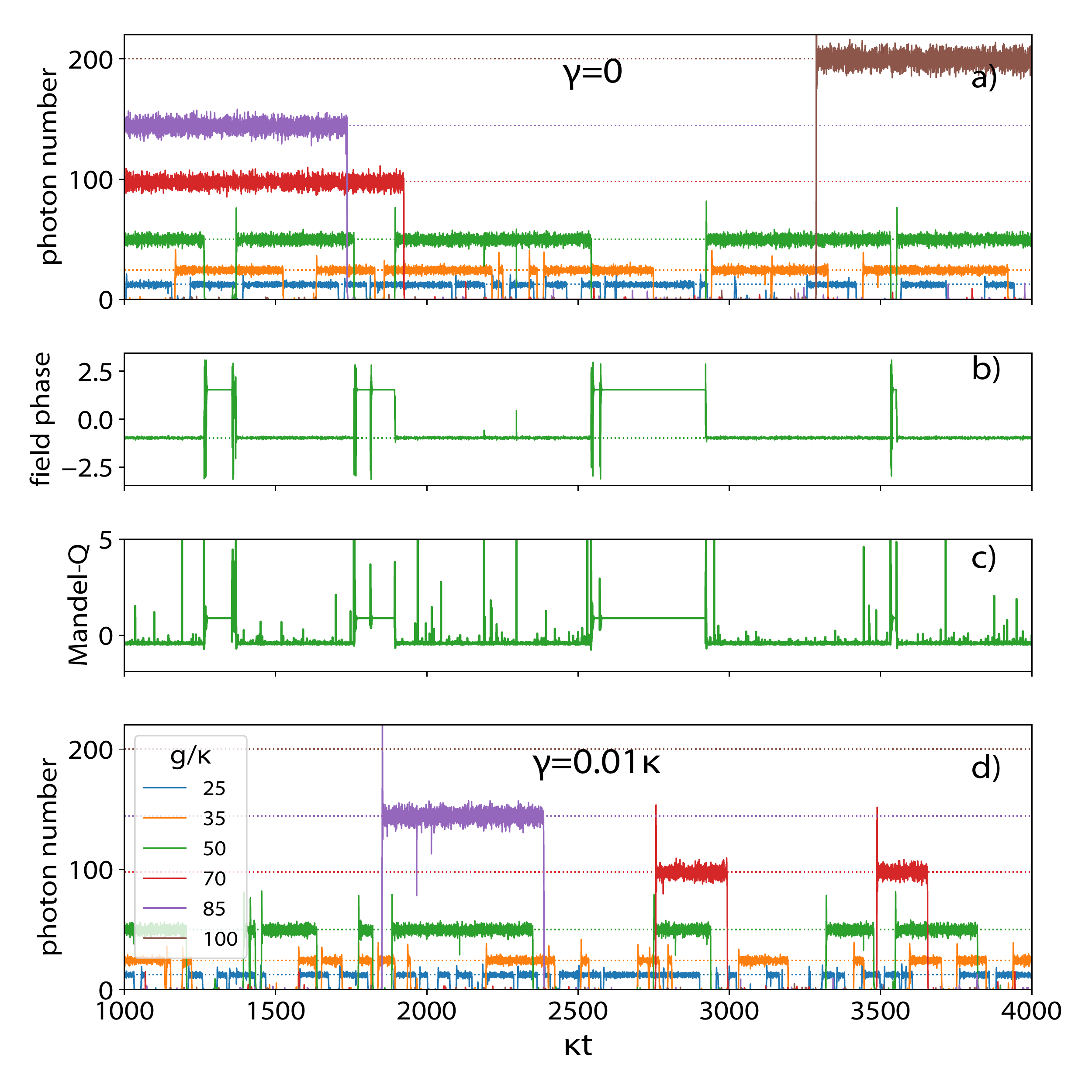}
 \caption{Example trajectories showing the blinking behaviour with different \(g\) values (color code is the same throughout). Parameters: \(\Delta=-5\kappa\), \(\eta=g/4\) (corresponding to the cyan star in \cref{fig:PhaseDiagram}). The dotted lines in panels a) and d) represent the estimate $g^2/(2\Delta^2)$ for the photon number, following from \cref{eq:resonance}. Panel b) shows the phase of the field along the green trajectory of panel a) ($g=50\kappa$), assuming a coherent state. The dotted green line is the field phase expected from \cref{eq:resonance}, that is, the phase of the complex number $\left(\kappa-i\left[\Delta+g/(2\sqrt{N})\right]\right)^{-1}$, substituting the bright-state photon number $N=50$ that can be read from panel a). We see that the coincidence with the simulated field phase in the ON periods is very good. Panel c) displays the Mandel-Q parameter along the same trajectory, exhibiting a larger nonclassicality for the OFF periods than the ON periods, which we will discuss in \cref{sec:DwellTimes}. Throughout this work, the numerics was performed with C++QED: a C++/Python framework for simulating open quantum dynamics \cite{VukicsCppQEDa,VukicsCppQEDb,Sandner2014CppQED}.}
 \label{fig:trajectories}
\end{figure*}

\begin{table*}
\centering
 \begin{tabular}{| r c l|}
 \hline
 ON value & $a$&\\ 
 blink-on rate & $\mu$& $=($ dwell time in OFF period$\;)^{-1}$\\
 blink-off rate & $\lambda$ & $=($ dwell time in ON period$\;)^{-1}$\\
 filling factor & $\mathcal F=\frac\mu{\mu+\lambda}$&\\
 expectation value & $\langle X\rangle=\frac{a\mu}{\mu+\lambda}$&\\
 variance & $\mathrm{var}\{X\}=\frac{a^2\mu\lambda}{(\mu+\lambda)^2}$&\\
 temporal correlation & $\langle X(t),X(t’)\rangle=e^{-(\mu+\lambda)\,|t-t’|}\;\mathrm{var}\{X\}$ &\\
 characteristic timescale & $\tau=(\mu+\lambda)^{-1}$&\\
 \hline
 \end{tabular}
 \caption{Characteristics of the telegraph process in the random variable $X$ in the special case when the OFF value is $0$.}
 \label{tab:telegraph}
\end{table*}

Obviously, due to the large difference in the photon numbers, the components of the mixture correspond to very distinct output signals. The photon number is continuously monitored via the photons outcoupled into the $\kappa$ loss channel. The classical distinguishability of large photocurrent versus dark counts amounts to a projection of the quantum state into only one of the components at a time. This is shown in \cref{fig:trajectories}, where the instantaneous photon number along quantum trajectories is plotted for various coupling strengths $g$. The bright and dark periods alternate sharply in the form of a telegraph signal. Detailed analysis of the statistical data shows that the presented signals are indeed very accurately (even to the limit of numerical accuracy) described by a telegraph process.\footnote{In particular, what we do is to define a binary signal from the somewhat noisy trajectories whose model is depicted in \cref{fig:trajectories}, simply by assigning the value 1 to the time instants where the photon number is higher than half of the temporal average and 0 to the others. On this binary signal $X(t)$ we verify the fulfillment of the relation $\mathrm{var}\{X\}=\langle X\rangle\left(1-\langle X\rangle\right)$, that is characteristic of the telegraph process (cf. \cref{tab:telegraph}). We find agreement up to \(10^{-14}\) precision.} This means that such trajectories have essentially three parameters: the amplitude of the bright period and the rates of blink-on and -off, $\mu$ and $\lambda$, respectively, since a telegraph process is nothing else than the composition of two temporal Poisson processes with exponential waiting-time distribution. Hence, the waiting time for a blink-on (the inverse of the blink-on rate $\mu$) equals the dwell time in the dim period, the same being true for blink-off and the bright period. The characteristics of the telegraph process are summarized in \cref{tab:telegraph}.

The trajectories for different coupling constants $g$ in \cref{fig:trajectories} are generated using the drive strength $\eta/g=1/4$ kept invariant, while fixing the detuning  $\Delta=-5 \kappa$. Hence all these curves correspond to the single point denoted with the cyan star in the phase diagram \cref{fig:PhaseDiagram} within the bistability domain. The photon number increases with increasing  \(g\). The numerics shows that the bright state has the photon number $N \approx g^2/(2\Delta^2)$, the corresponding dotted straight lines fitting nicely on the noisy numerical record. This $N$ is twice the photon number at the lower boundary of the bistability range of the neoclassical theory. Interestingly, this analytical estimate satisfies the simple classical relation \labelcref{eq:resonance} with high accuracy, which suggests that the intuitive picture of near-resonantly driving an equidistant ladder holds even at such drive strength far above the lower boundary of the bistability domain. The same is suggested by \cref{fig:trajectories}(b), where the field phase is plotted together with an estimate from the same \cref{eq:resonance}, once more to yield excellent agreement (cf. the figure caption for details).

Since $N_\mathrm{scale} \propto g^2$, the photon number measured in units of $N_\mathrm{scale}$ proves to be invariant for the different curves. On the other hand, the dwell times in the attractor states increase significantly with increasing $g$. This reveals that there is a thermodynamic limit in which the phases become robust, the telegraph signal would disappear and it would be replaced by a hysteresis-like behaviour in a genuine first-order phase transition.


\Cref{fig:trajectories}(d) shows that the addition of a small amount of atomic decay (\(\gamma=0.01\kappa\)) does not lead to a qualitative change of the conclusions above. The photon number of the ON period remains the same as with \(\gamma=0\), however, atomic decay does noticeably decrease the dwell time in this attractor.

\section{Filling factor and the scaling of the drive strength in the thermodynamic limit}
\label{sec:fillingFactor}

\begin{figure}
\begin{center}
 \includegraphics[width=\linewidth]{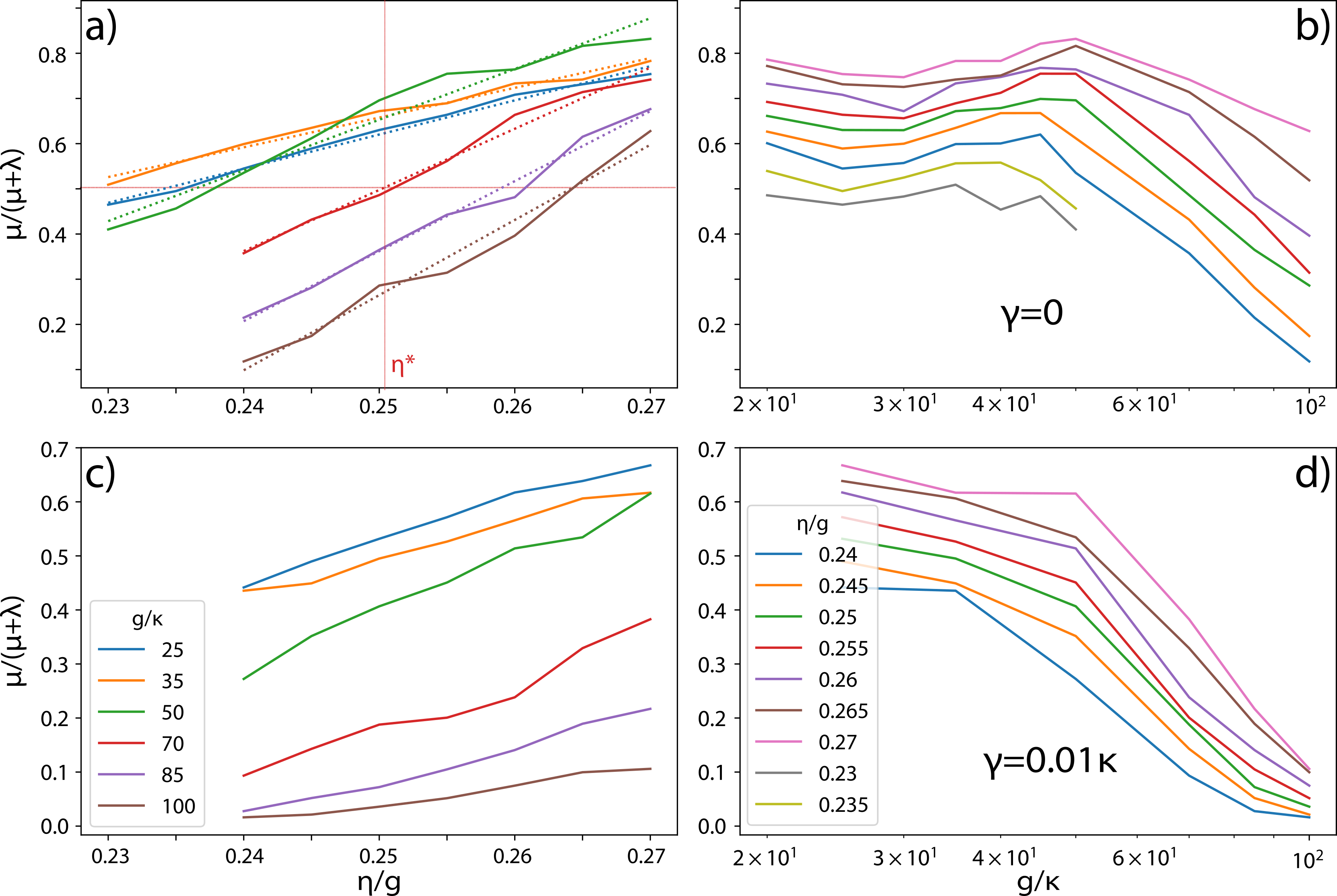}
\end{center}
 \caption{Filling factor of the bright periods (cf. \cref{tab:telegraph}) as a function of \(\eta/g\) and \(g/\kappa\) without (panels a and b) and with a small (panels c and d) atomic decay $\gamma=0.01\kappa$. Color code is the same in the upper and lower row.
 Panel a): the dotted lines indicate linear fits in order to determine $\eta^*$, the drive strength leading to half filling. The value of $\eta^*$ is indicated for the coupling strength \(g=70\kappa\). }
 \label{fig:FillingFactor}
\end{figure}

In the bimodal steady-state density operator, the weights of the components vary through the bistability domain, in particular, on increasing the drive strength the ‘dim’ state component vanishes gradually in favour of the ‘bright’ state. In the time domain, the telegraph signal manifests this form of transition through the variation of the filling factor which is theoretically  $\mathcal F = \frac{\mu}{\mu+\lambda}$. The trajectories in \cref{fig:trajectories} show that the filling factor varies for the telegraph signals with different $g$.  This is confirmed in \cref{fig:FillingFactor}. Panel (a) shows the monotonous increase of $\mathcal F$ as a function of $\eta/g$ for a set of $g$ values. The curves are shifted with respect to each other. This dependence is made explicit in panel (b), where $g$ is varied while keeping $\eta/g$ at various fixed values. The filling factor is constant in the range of smaller $g$ values, whereas there is a decrease of $\mathcal F$ in the range of larger coupling strengths. For example, the green line represents a closely constant filling factor around $2/3$ up to $g\approx 50 \kappa$, but if $g$ is increased further,  the filling factor drops.\footnote{The reason why the curves span different ranges in panels (a) and (b) is that we needed to use different ranges of \(\eta\) for different \(g\) values in order to find the value of \(\eta^*\).}

It follows then that the scaling of the drive strength $\eta$ such that $\eta/g$ is kept invariant does not preserve the self-similarity of the telegraph signal. Scaling to the thermodynamic limit means that the system keeps a self-similar behaviour, only scaled up in time and brightness of the ON state. Since the latter two scale with the single parameter $g$, there is the parameter $\eta$ left at our disposal to ensure self-similarity during the upscale, which, in the case of such a simple process as the telegraph one, cannot mean anything else than keeping the filling factor constant. As the most obvious choice, we pick the case when the filling factor is 0.5.

Since the concept of self-similarity is a difficult one in the present context, let us employ a more pictorial explanation. Upscaling means that we are looking on the system’s time evolution through a telescope that via the turn of a \emph{single} knob (here, increasing \(g\), the single scaling parameter), increases its angle of view both in time and photon number,\footnote{The increase in these two dimensions is not necessarily at the same rate, e.g. here the the photon number increases under the fairly obvious rule \(g^2\), while the timescale as \(g^\nu\), where \(\nu\) is one of the finite-size scaling exponents that we want to find in this study.} but keeps projecting the image on the same ocular area. (Hence, it has an increasingly coarse resolution both in time and photon number.) This is like the “zoom” functionality on modern camera objectives. It is very important that we are aiming at a single-parameter scaling theory, that is, we have to find a rule for how to change the other parameters of the system (here, only \(\eta\)) as a function of the scaling parameter \(g\) during the upscale, i.e. how to “scale \(\eta\) with \(g\)”. The rule that defines such a one-dimensional manifold in the parameter space as the upscale orbit is: \emph{self-similarity}. Self-similarity means that we require the image of the system’s time evolution on the ocular of the telescope to remain the same during this procedure. In the case of such a simple process as the telegraph signal, the filling factor is the single parameter that determines the image in such a telescope. Hence, self-similarity means that we are keeping the filling factor constant, namely 0.5.

\begin{figure}
 \centering \includegraphics[width=\linewidth]{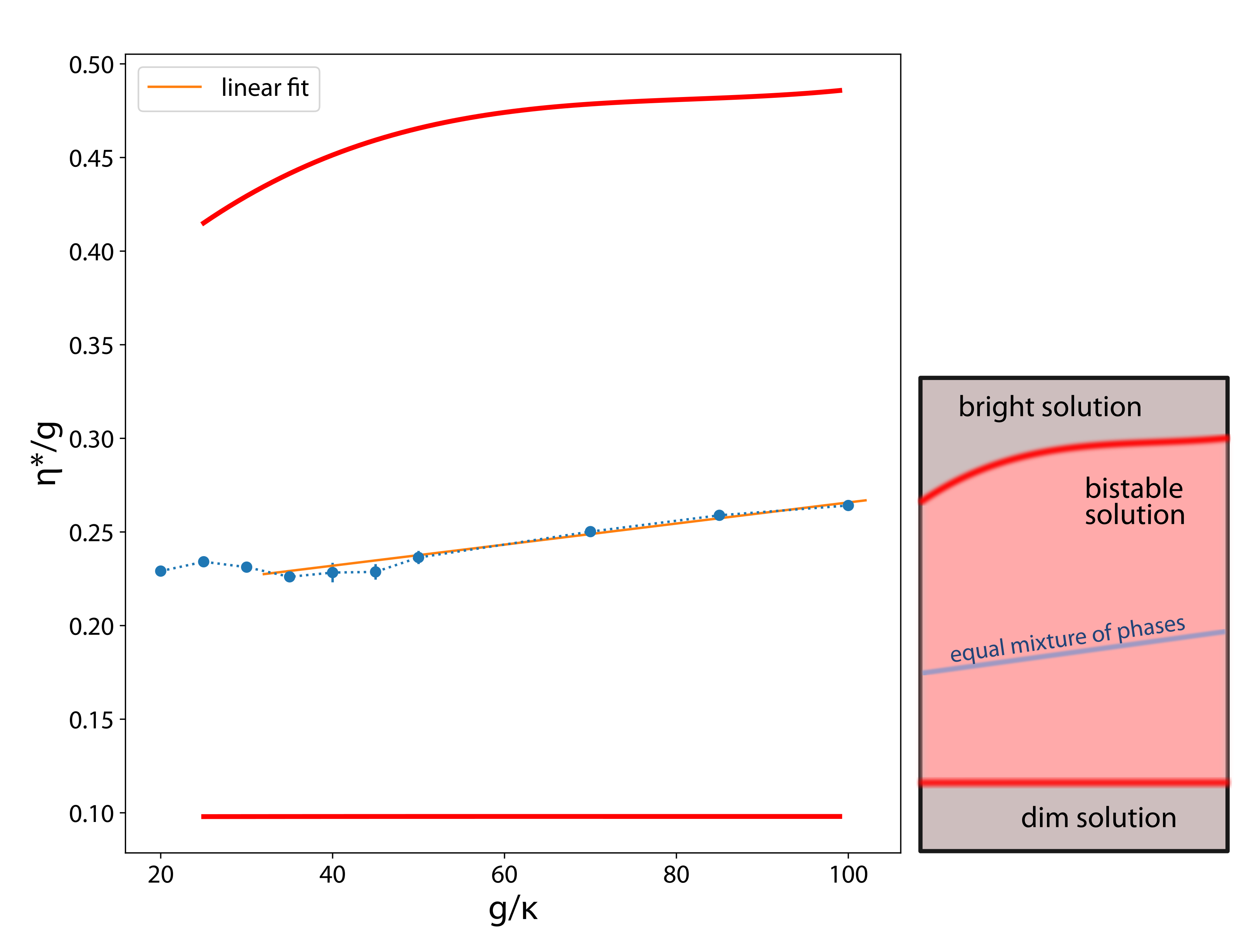}
 \caption{The drive strength \(\eta^*\) leading to half filling of the telegraph signal as a function of \(g\), i.e., the equal mixture of the two phases in the steady-state density matrix. Solid red curves represent the phase boundaries of the bistability domain from the classical theory.}
 \label{fig:HalfFilling}
\end{figure}

Therefore, in order to determine the correct scaling of $\eta$ in the thermodynamic limit, we have to find the drive strength for each $g$ value (denoted $\eta^*(g)$ in the following) that leads to half filling. Looking at \cref{fig:FillingFactor}(a), we observe that an approximate linear interpolation (linear fit) captures appropriately the behaviour of the $\mathcal F(\eta)$ curves for the different $g$ values in the range of interest. Hence, we use such an interpolation to determine $\eta^*(g)$. As an example, in the plot we show for the red curve ($g=70\kappa$) the corresponding $\eta^*$. In \cref{fig:HalfFilling}, we plot the function $\eta^*(g)$, embedded in the bistability domain that we have calculated similarly to \cref{fig:PhaseDiagram}. With this numerical result \emph{we defined the appropriate finite size scaling}.  As it turns out, a linear fit reproduces quite exactly the behaviour of $\eta^*$ for $g\gtrsim50\kappa$, which indicates that the correct scaling of the drive strength in the thermodynamic limit is \(\eta\sim g^2\).\footnote{It is conceivable that had we included \(\Delta\) in the scaling as suggested in Ref. \cite{gutierrez2018dissipative} (\(\Delta/g=\mathrm{const.}\)), then the \(\eta/g=\mathrm{const.}\) scaling would have been sufficient to preserve self-similarity. However, to answer this question rigorously, we would need to repeat the whole numerical work for that scaling also, that is beyond the scope of the present study.}

\begin{figure*}
\centering \includegraphics[width=\linewidth]{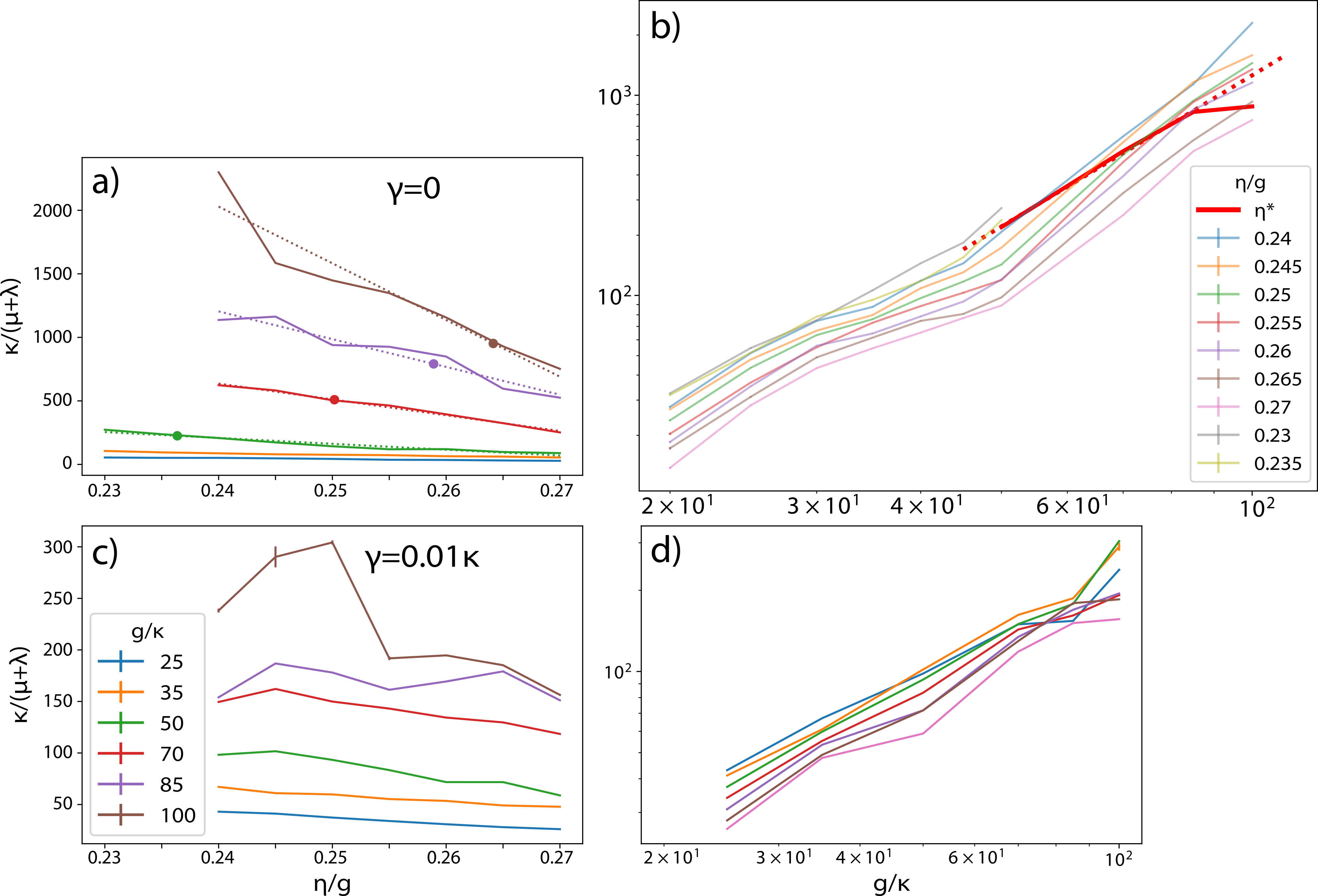}
 \caption{The characteristic timescale of the bistable blinking process given by the inverse of $\mu+\lambda$. The points in panel a) correspond to the values of \(\eta^*\) for the different \(g\) values where half-filling of the telegraph signal is achieved. In panel b), the solid red curve designates the timescale at \(\eta^*\) as a function of $g$, that is, the timescale in the correct finite-size scaling when the telegraph process is kept at half filling during the passage to the thermodynamic limit. The dotted line shows a log-log linear fit on the values leaving out \(g=100\kappa\). The exponent resulting from the fit is roughly 2.2. Panels c) and d): timescale with finite atomic decay $\gamma=0.01\kappa$.}
 \label{fig:dwellTime}
\end{figure*}

\section{Dwell times}
\label{sec:DwellTimes}

\subsection{Characteristic timescale of the telegraph signal}

Figure \ref{fig:dwellTime} presents the characteristic timescale of the bistable blinking process. This is defined as the inverse of the sum of blinking rates $\mu+\lambda$, which is extracted from an exponential fit on the numerically calculated temporal self-correlation of the signal (cf. \cref{tab:telegraph}).
We use the unit of $\kappa$ so that the large difference is manifested in the figure: the characteristic times are orders of magnitude above the microscopic timescale  $\kappa^{-1}$. On increasing the drive strength $\eta$ in the presented range, the average dwell time decreases because of the increase of the rate $\mu$ of blink-on, in conjunction with the increase of the filling factor. Points represent the values of $\eta^*$ where the telegraph signal has half-filling.

From the point of view of the thermodynamic limit, the dependence of the characteristic timescale on $g$ is the most relevant (\cref{fig:dwellTime}(b)). We show the increasing timescale over two orders of magnitude of the coupling constant $g$ for various values of $\eta/g$. This $g$ range is given by computational limitations, nevertheless, it is enough to demonstrate \emph{the power-law scaling} of the increase of the timescale and to determine the exponent. To obey the correct self-similar scaling detailed in \cref{sec:fillingFactor}, we have to find the timescale for the drive strength $\eta^*(g)$ for the different $g$ values. To this end, we again use a linear fit on the curves of $\eta$-dependence, which captures the behavior quite correctly (cf. dotted lines in the panel (a), with the $\eta^*$ values depicted with big dots of the corresponding colour). The timescale change under finite-size scaling $\tau\left(g,\eta^*(g)\right)$ is shown in the thick, solid, red line in \cref{fig:dwellTime}(b). A linear fit in the log-log scale leads to the numerical estimate of 2.2 for the finite-size scaling exponent of the characteristic time.
The point $g=100\kappa$ was omitted from the fitting because the dwell times were systematically underestimated due to truncation of the trajectories for the long but finite simulation time.


\begin{figure}
 \includegraphics[width=\linewidth]{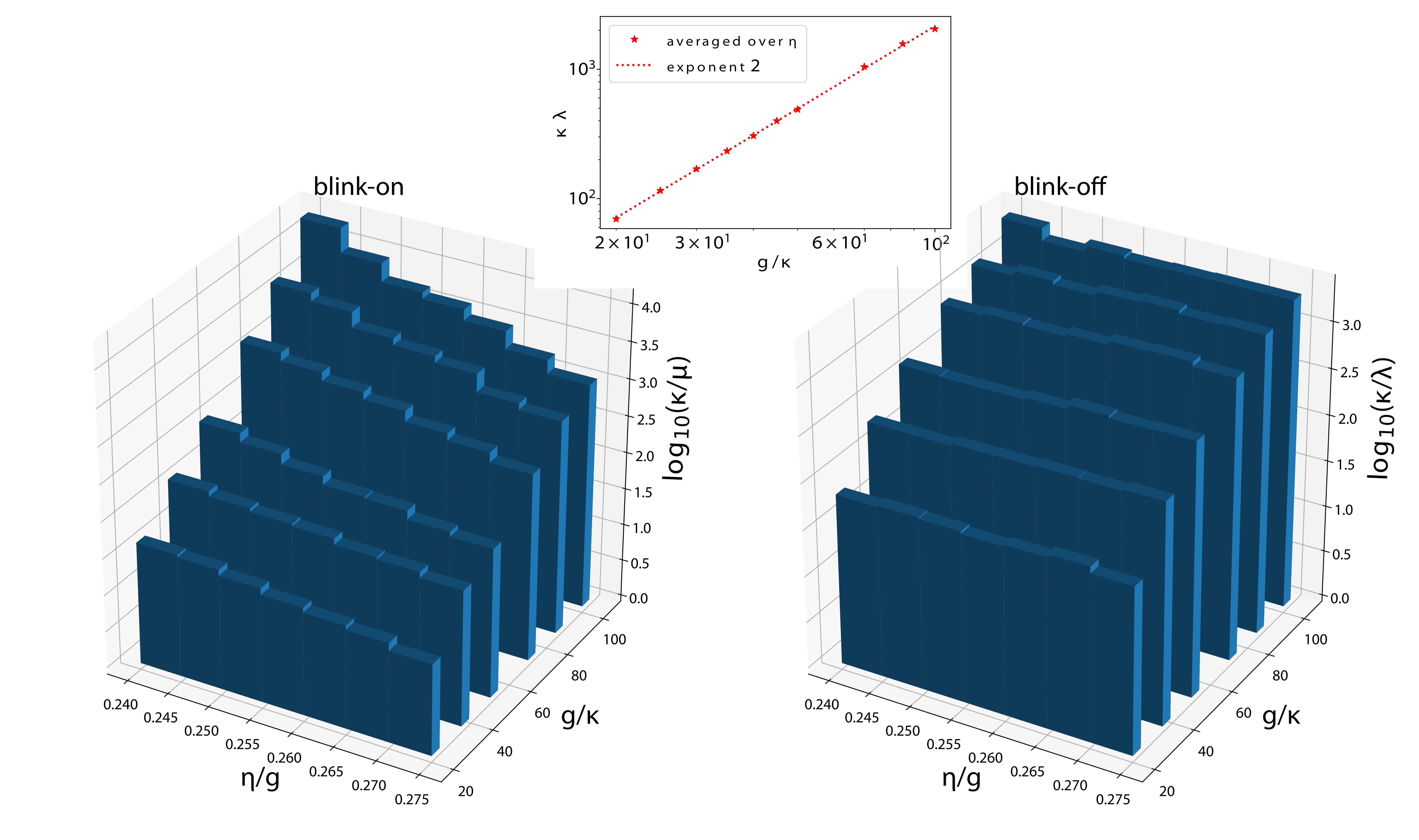}
 \caption{Waiting time for blink-on (\(\mu^{-1}\)) and blink-off (\(\lambda^{-1}\)). Inset: $\lambda$ values averaged over $\eta$ as a function of $g$, with an exponential fit with exponent 2.}
  \label{fig:Rates}
\end{figure}


\Cref{fig:Rates} is devoted to the numerical analysis of the rates of blink-on and -off, $\mu$ and $\lambda$, respectively, which sheds light on the physical processes leading to switching between the robust classical attractor states in the finite-size system. These are calculated from combining the above-discussed characteristic timescale with the filling factor. For both rates, the increase of $g$ implies a reduction, in agreement with the expectedly growing stability of phases on approaching the thermodynamic limit.

\subsection{Blink-off process}

The downward process is initiated by a single photon loss (detection) event. This is because there is a chance that the state residing in the ladder $\ket{n,-}$ gets projected into the ladder $\ket{n,+}$ under such an event \cite{carmichael2015}, since 
\begin{eqnarray}
 a\ket{n,\pm}=&\frac{\sqrt n+\sqrt{n-1}}2\ket{n-1,\pm}+\frac{\sqrt n-\sqrt{n-1}}2\ket{n-1,\mp}\nonumber\\&\approx\sqrt n\ket{n-1,\pm}+\frac{1}{4\sqrt n}\ket{n-1,\mp}.
\label{eq:ladderSwitching}
\end{eqnarray}
Once such a jump occurs, there is a downward cascade of photon escapes, because while our (red-detuned) drive was closely resonant with the high-lying part of the $\ket{n,-}$ ladder, resulting in an approximately coherent state in this part, it is off-resonant with the $\ket{n,+}$ ladder (which would be resonant with the drive blue-detuned with the same amount), so on this ladder there is no drive to compensate the photon loss. The passage downward consists of a quick cascade of many jumps amounting to an exponential decay.

Although a single ladder-switching quantum jump initiates the blink-off, however,  the likelihood of such a rare jump vanishes completely in the $g, N \rightarrow\infty$ thermodynamic limit.  The rate of ladder switching scales as $\kappa/N$ (from Eq.~(\ref{eq:ladderSwitching}), the total jump rate scales as $\kappa N$, while the probability of a ladder switch to occur within a jump is $\propto1/N^2$), and as we saw above, the bright-state photon number scales as $g^2$ in the thermodynamic limit. This would suggest the exponent 2 for the finite-size scaling of the blink-off timescale, which is verified in the inset of the right panel of \cref{fig:Rates} to very good accuracy. This downward jump rate $\lambda$ is largely independent of $\eta$, that only slightly influences the photon number in the bright phase. 

\subsection{Blink-on process}

On the other hand, the switching from the dim to the bright phase is induced by the external driving and thus is sensitive to $\eta$, as can be seen in the left panel of \cref{fig:Rates}. The upward process is suppressed by the off-resonance of low-lying quasi-energy levels (anharmonic part of the spectrum). Therefore, it is easy to understand that the larger the coupling $g$, the larger the shift of the levels from resonance and the smaller the blink-on rate. In the dim phase, the state of the bosonic mode is close to the vacuum, however, there must be a small deviation from that due to the driving. The state is a non-classical superposition with positive Mandel-Q parameter. The Mandel-Q parameter measuring the nonclassicality of the field state is defined as
\begin{equation}
 Q=\frac{\mathrm{var}(a^\dagger a)-\langle a^\dagger a\rangle}{\langle a^\dagger a\rangle},
\end{equation}
where the averaging can be performed either as a quantum average on the actual state vector of the field to reflect the nonclassicality at a given time instant, or also over time. The Q parameter is zero for a classical field state (coherent state). In \cref{fig:trajectories}(c), $Q$ is calculated as a time-dependent quantum average, and we observe that the nonclassicality is stronger in the dim phase than in the bright one. This is consistent with our picture that the bright phase consists of an approximately coherent state on the manifold $\ket{n,-}$.

The state in the dim period has the property that the projection of the wavefunction after a photon detection increases the weight of a high-excitation component. It can be shown that for pure states of a mode, the positivity of the Mandel-Q parameter, which is what we have in the dim phase, is equivalent to the nonclassical situation that a photon escape from the mode \emph{increases} its photon number. While in the dim state there is a negligible amount of excited photon component generated by the $\eta$ driving, triggered by a single quantum jump (which is very rare on account of the very low dim-state photon number), it can grow in an exponential runaway process for subsequent photon detections. So the blink-on also takes place in the form of a cascade of quantum jumps. 

\begin{figure}
 \includegraphics[width=\linewidth]{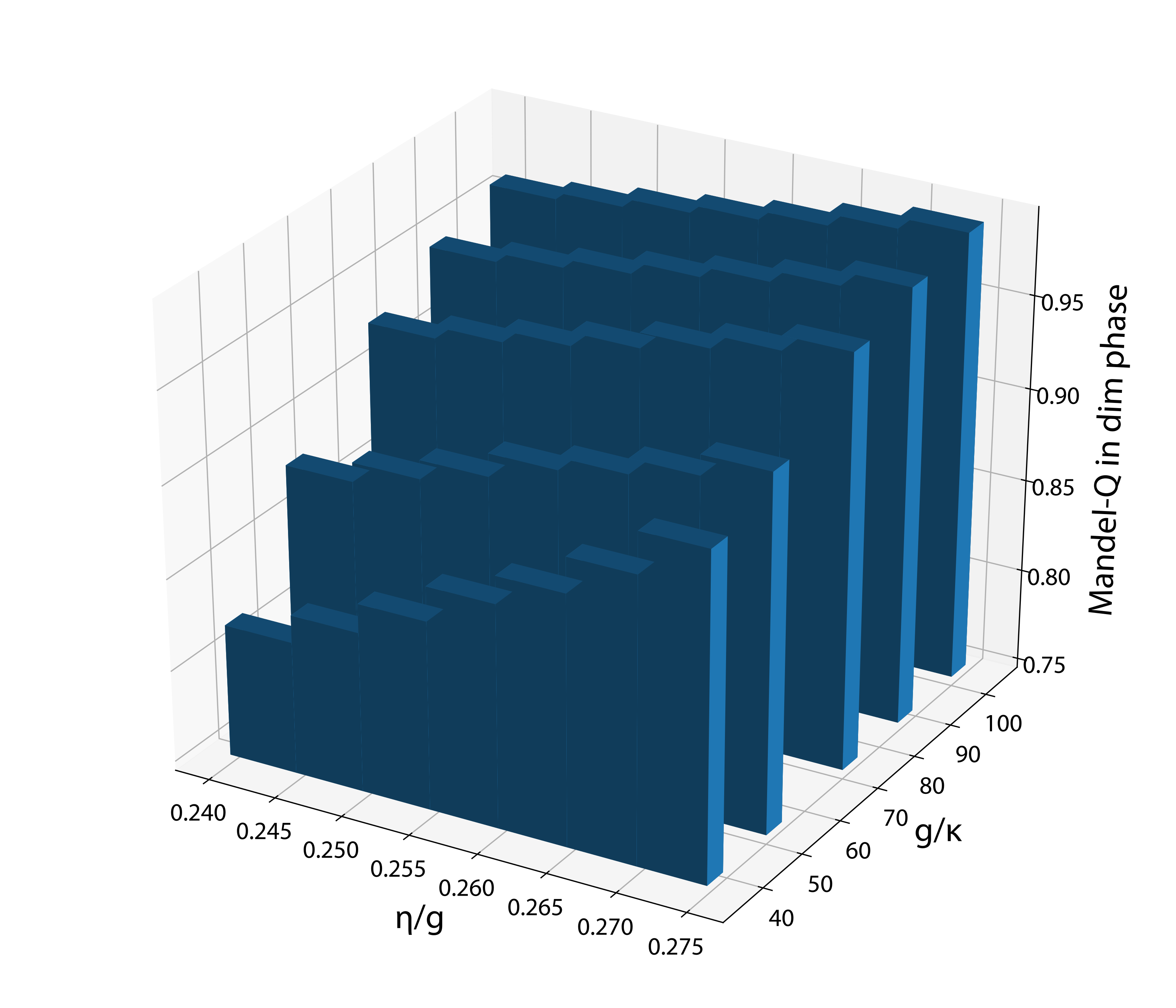}
 \caption{Mandel-Q parameter of the field averaged over time in the dim periods.}
  \label{fig:MandelQ}
\end{figure}

In \cref{fig:MandelQ}, we plot the field Q parameter averaged over time, but only in the dim periods.\footnote{Note that a time average of the Mandel-Q calculated for the instantaneous pure states of trajectories does not reproduce the Q of the time-averaged density operator, since the dependence of Q on the density operator is not linear. The aim of our usage is to substantiate that the quantum state of the mode in the dim periods is of the form \labelcref{eq:inferredFieldDimState}, where \emph{a photon escape increases the photon number of the mode}.} The overall trend is that the nonclassicality of the field in the dim phase increases both with increasing $\eta$ and $g$, and the dependence flattens out for large $g$ at a value close to 1. Hence, the dim phase remains nonclassical also in the thermodynamic limit.

Let us try to model the dim state of the field in the following form:
\begin{equation}
\label{eq:inferredFieldDimState}
 \ket{\Psi}=\sqrt{1-\varepsilon^2}\ket{0}+\varepsilon\ket{\varphi},
\end{equation}
where $\varepsilon$ is a small number and $\ket{\varphi}$ is a state orthogonal to the vacuum state. The photon count rate  $\kappa\langle a^\dagger a\rangle = \kappa\,\varepsilon^2 \, N_{\ket{\varphi}}$ can be made very small with  $\varepsilon\rightarrow 0$, where $N_\varphi = \langle\varphi| a^\dagger a| \varphi \rangle$ is the mean photon number associated with the component $\ket{\varphi}$ superposed on the vacuum. At the same time, the Mandel-$Q$ parameter of the state \labelcref{eq:inferredFieldDimState} is found to be independent of $\varepsilon$ in the limit $\varepsilon\ll1$: one finds $Q_{\ket{\Psi}}= Q_{\ket{\varphi}} + N_{\ket{\varphi}}$. That is, it depends only on the properties of the state $\ket{\varphi}$. Note that (i) $N_{\ket\varphi} > 1$ because the vacuum component is missing from photon number expansion of  $\ket{\varphi}$, and (ii) the Mandel-$Q$ parameter is always limited below 1 in \cref{fig:MandelQ}. These two observations imply  that the state $\ket{\varphi}$ is a nonclassical state with negative Mandel-$Q$ factor. 

\begin{figure*}
\begin{tabular}{l l l}
(a) & (b) & (c) \\
\includegraphics[width=.31\linewidth]{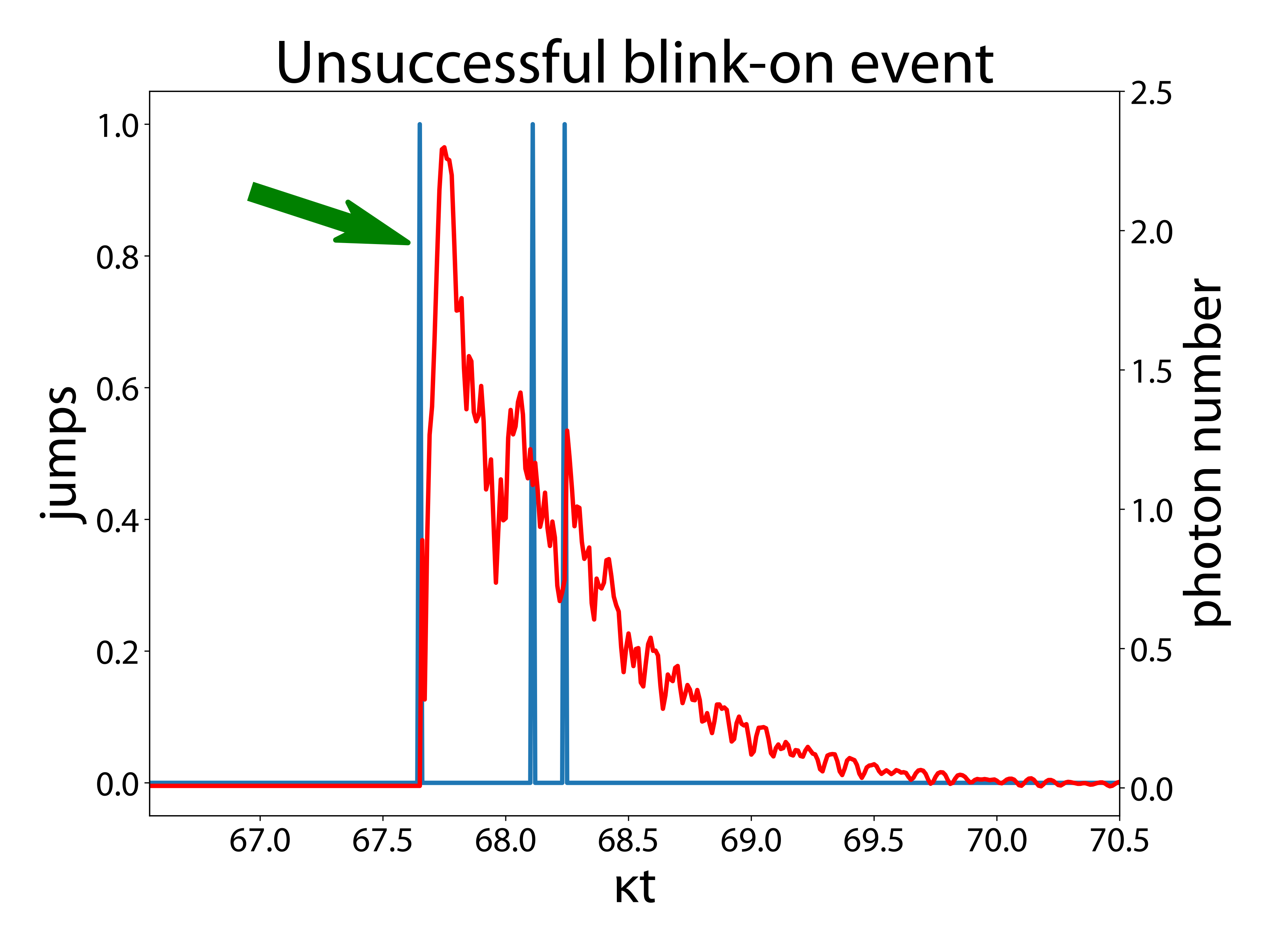} & \includegraphics[width=.31\linewidth]{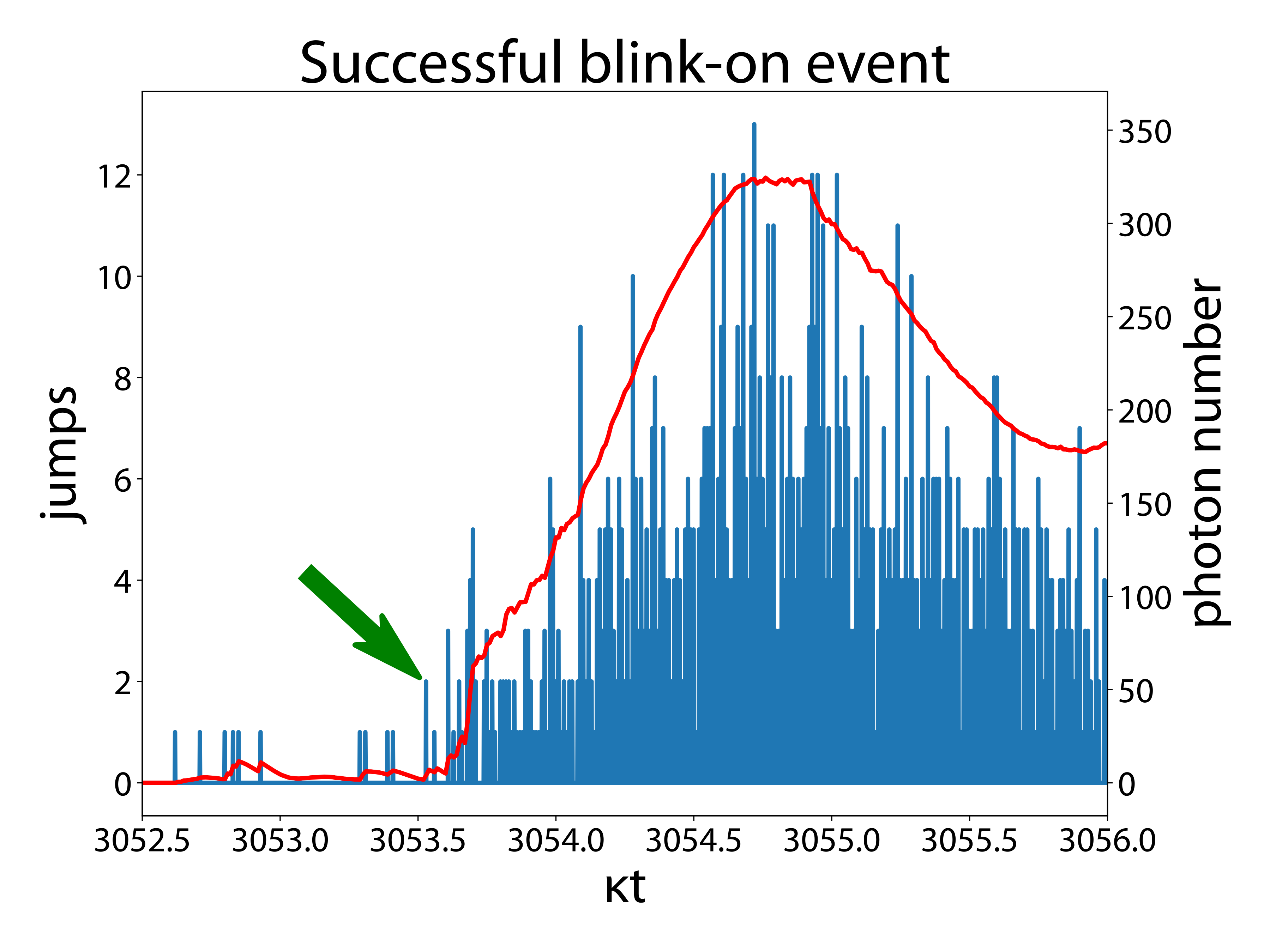} & \includegraphics[width=.31\linewidth]{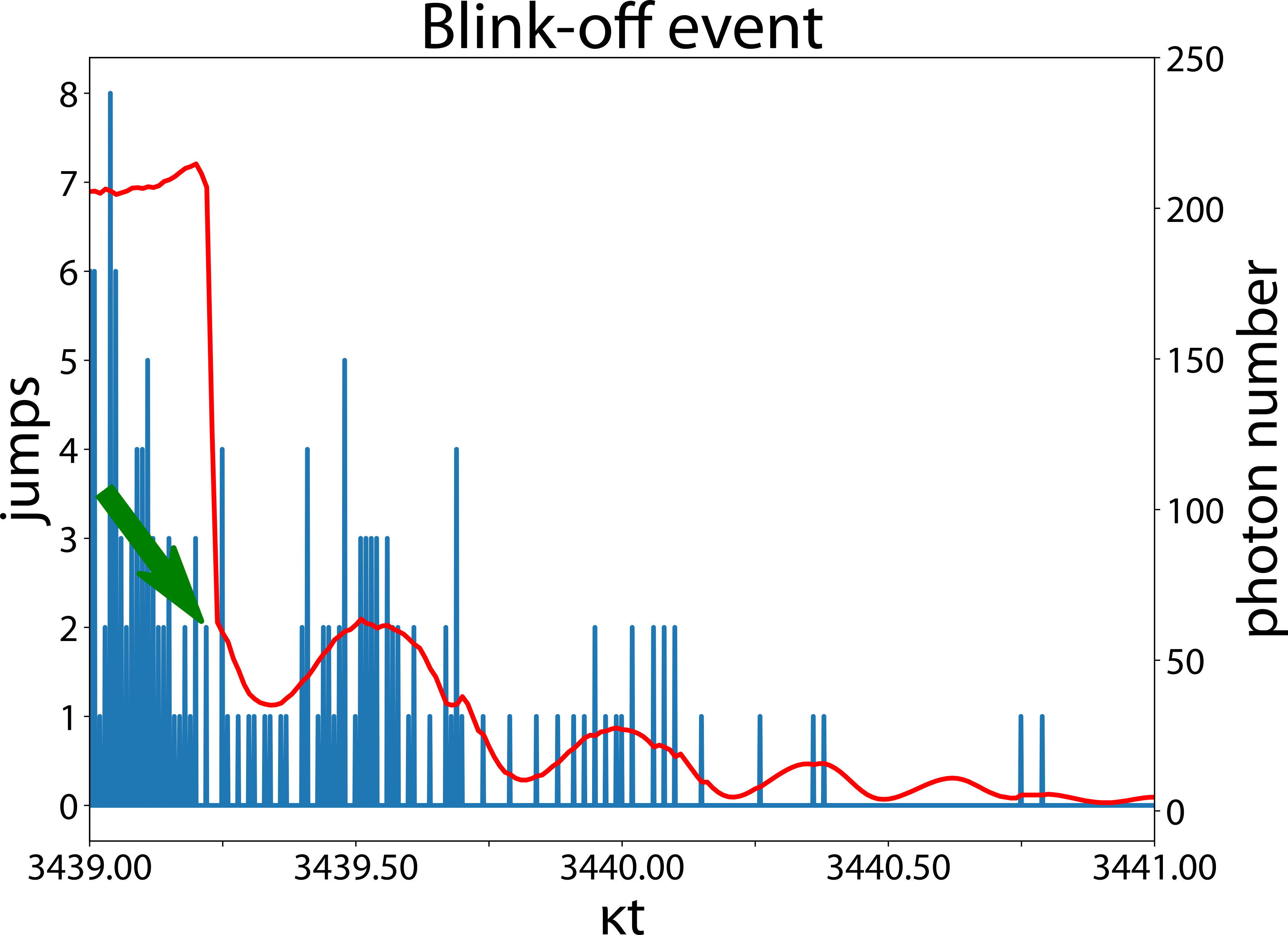}
\end{tabular}
 \caption{Photon-escape quantum jumps (blue) and the evolution of instantaneous photon-number expectation values (red) during blink-on and -off events. Green arrows indicate the photon escapes that trigger the switching events. The time-resolved depictions of quantum jumps are histograms with bin sizes \(0.01/\kappa\) for (a) and \(0.001/\kappa\) for (b,c).}
 \label{fig:shootupdown}
\end{figure*}

\subsection{Cascades of quantum jumps switch between attractors}

A study of the photon-number evolution along a single trajectory together with sufficiently resolved quantum jump events confirms the above picture both for the blink-on and -off events, cf. \cref{fig:shootupdown}. In panel (a) we see that a single photon escape (marked with the green arrow) triggers a shootup of the photon number from the dim state, in accordance with the dim state being nonclassical with positive Mandel-Q parameter. However, in this event the surge is not strong enough to break the blockade. Panel (b) depicts a successful breakthrough event where we also see that the buildup of the full bright-state photon number incurs a \emph{probabilistic cascade of quantum jump events}. In panel (c) also a single quantum jump triggers the collapse of the bright state (green arrow at the sudden drop of photon number), followed by a normal ringdown of the photon number with rate \(\kappa\), involving several further photon escapes. This proves our claim made in the Introduction that the telegraph signal observed here differs essentially from the electron-shelving scheme: though there is a trigger single-photon escape (quantum jump), the switch between the dim and bright phases is driven by a cascade of quantum jumps.

\section{The role of atomic decay}
\label{sec:AtomicDecay}
The condition $\gamma=0$ is essential in the neoclassical theory, since the derivation of the transcendental \cref{eq:nscale} relies heavily on the fact that the length of atomic pseudo-spin is conserved, $\langle \sigma_x \rangle^2 + \langle \sigma_y \rangle^2 + \langle \sigma_z \rangle^2 = 3/4$. Allowing $\gamma\neq0$ hence leads to qualitatively different behaviour since this conservation law is broken.

Yet, for very small $\gamma$ values, the behavior of the full quantum model does not seem to be qualitatively affected: as we have shown in the course of this study in passing, the photon-blockade-breakdown effect can be observed in the case of a finite but small $\gamma$.

The effect of an atomic decay in the bright state, i.e., on a high-lying coherent state can be assessed similarly to the above:
\begin{equation}
 \sigma\ket{n,\pm}=\frac1{\sqrt{2}}\left(\ket{n-1,\pm}+\ket{n-1,\mp}\right).
\end{equation}
This means that in the event of an atomic decay, there is a $1/2$ probability of a ladder switch. Hence, a $\gamma$ on the order of $\kappa$ would wipe out the blinking effect that is the central theme of this paper, since a blink-on would immediately be followed by a collapse of the bright state. The blinking effect manifests itself most clearly when $\gamma=0$, the case that we considered mostly here.

If $\gamma\ll\kappa$, which case we exposed in passing in \cref{fig:trajectories,fig:FillingFactor,fig:dwellTime}, the atomic decay emerges as a competing timescale for high-enough bright-state photon numbers, resulting overall in smaller characteristic times and filling factors. With respect to the phase transition, the system is interesting as long as the microscopic timescale of spontaneous emission is longer and thus is dominated by the shorter macroscopic timescale $\tau$ of the phase stability.

\section{Conclusions}
\label{sec:conclusion}

The time evolution of a quantum system undergoing a dissipative, first-order quantum phase transition has been studied. We considered the photon-blockade-breakdown phase transition, which takes place in the driven-dissipative Jaynes-Cummings model with strong coupling between the two-level system and the harmonic oscillator. For a certain range of drive strength, the stationary solution corresponds to a bistability of classically distinguishable states. By unraveling the stationary solution into quantum trajectories, we resolved the nature of coexistence of phases. We constructed an appropriate scaling of the system parameters such that the bistability is manifested by a self-similar telegraph signal. The finite-size scaling argument verifies that the bistability solution develops into a first-order phase transition in the thermodynamic limit. We calculated the finite-size scaling exponents numerically. Even in the thermodynamic limit, the stability of phases originates from the discrete spectrum of a small quantum system and the dim phase exhibits nonclassical photon statistics.

\section*{Acknowledgements}

On behalf of Project WAL, we thank for the usage of MTA Cloud (https://cloud.mta.hu/) that significantly helped us achieving the results published in this paper. This work was supported by the National Research, Development and Innovation Office of Hungary (NKFIH) within the Quantum Technology National Excellence Program  (Project No. 2017-1.2.1-NKP-2017-00001) and by Grant No. K115624. A. Vukics acknowledges support from the János Bolyai Research Scholarship of the Hungarian Academy of Sciences.

\bibliography{references}

\end{document}